\documentclass[twocolumn,10pt]{article}
\usepackage{aaai}
\usepackage{times}
\usepackage{helvet}
\usepackage{courier}
\usepackage[utf8]{inputenc} 
\usepackage[T1]{fontenc}    
\usepackage{hyperref}       
\usepackage{url,multirow}            
\usepackage{booktabs}       
\usepackage{amsfonts}       
\usepackage{nicefrac}       
\usepackage{microtype}      
\usepackage{lipsum,graphicx,amsmath}
\graphicspath{{figs/}}
\usepackage{algorithmic,algorithm}
\usepackage{amsmath}

\hypersetup{draft}
 
\begin{document}

\title{Humans learn too: Better Human-AI Interaction using Optimized Human Inputs} 

\author{Johannes Schneider\\
$^1$University of Liechtenstein, Vaduz, Liechtenstein\\
johannes.schneider@uni.li
}

\maketitle %
\begin{abstract} 
Humans rely more and more on systems with AI components. The AI community typically treats human inputs as a given and optimizes AI models only. This thinking is one-sided and it neglects the fact that humans can learn, too. In this work, human inputs are optimized for better interaction with an AI model while keeping the model fixed. The optimized inputs are accompanied by instructions on how to create them. They allow humans to save time and cut on errors, while keeping required changes to original inputs limited. We propose continuous and discrete optimization methods modifying samples in an iterative fashion.  Our quantitative and qualitative evaluation including a human study on different hand-generated inputs shows that the generated proposals lead to lower error rates, require less effort to create and differ only modestly from the original samples.

\end{abstract}

\section{Introduction}
In an industrial setting, production processes involving humans and robots are constantly improved to reduce errors and improving efficiency. Improvements target measures towards humans and machines alike in a holistic manner. In the machine learning community, the setup is usually different: Treat the dataset originating from people as fixed and find the best possible model.
\begin{figure}
	\centering \includegraphics[width=0.45\textwidth]{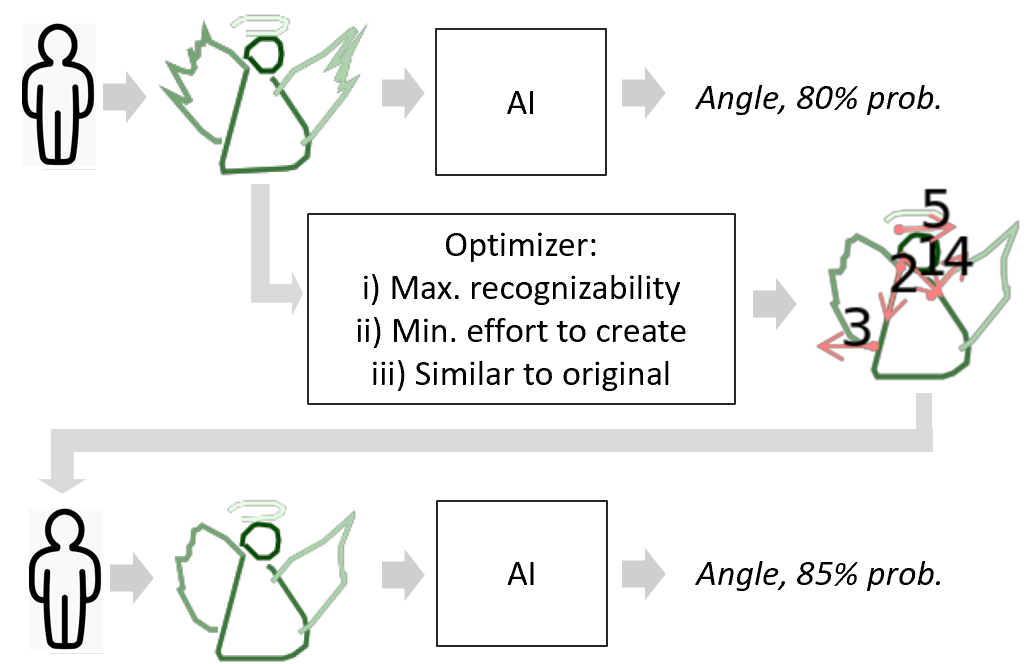}
	\caption{Humans change their inputs to AI based on feedback leading to time savings and better recognizability} \label{fig:over}
	\vspace{4pt}
\end{figure}

This works breaks with this paradigm. It focuses on helping humans to improve, i.e. humans are offered feedback allowing them to alter their inputs to machine learning systems. See Figure \ref{fig:over} for a process overview. By improving their interaction with AI systems, humans might benefit from lower time to create inputs and fewer misunderstandings in interaction.  
This might be highly attractive, since ``smart systems'' containing machine learning components have already deeply penetrated our daily life and humans interact with multiple such systems on a daily basis: recommendation systems (on web pages), voice assistants, gesture recognition systems, driver-assisted cars, to name a few. Furthermore, thanks to the recent success of AI and related technologies such as deep learning, interaction is likely to continue and include more and more safety-critical systems, such as fully self-driving cars, where errors in interaction might have fatal consequences. Thus, there are good reasons, why humans might want to improve on their inputs. Furthermore, if human inputs are of very low quality, so that they are either not recognizable or easy to confuse, even highly sophisticated machine learning systems might not be able to correctly interpret them. These points illustrate the need to help humans improve on their behavior.
Humans should be educated to produce inputs that reduce errors in interaction and require little effort to create. Furthermore, learning the proposed changes should also be easy. Unfortunately, fulfilling all of these objectives is difficult for multiple reasons. First, there is a trade-off between accuracy in recognition and effort to create. Simplifying inputs too much will increase error rates. Making inputs more complex, e.g. by adding redundant features to inputs, might reduce errors, but increase the amount of effort to create them.  Second, streamlining human-to-AI interaction is more intricate than human-to-human interaction, since AI systems process information differently and are sensitive to input changes that are hardly noticed by humans. Using almost invisible, adversarial perturbations, classifiers might be ``fooled'' to misjudge samples that can be categorized without any problems by humans (e.g. \cite{pou18}). From our perspective this is both good and bad news. While it suggests that minor changes exist that might increase confidence in the correct class (rather than in an incorrect class as used in an adversarial setting), it also highlights that humans might not even be able to distinguish an improved from an non-improved sample if optimization is not done with care. Moreover, human controlled movements of limbs (and the vocal tract) is subject to stochastic variation, making it close to impossible to alter inputs in very subtle ways as done in adversarial settings like \cite{pou18}. \\ 
Designing adequate optimization objectives is non-trivial. Even if the proposed samples optimize a specific mathematical objective, it is not clear, how well humans can actually learn such changes. That is, humans might deem the proposed inputs unnatural and they might, at least initially, struggle to reproduce such changes that deviate from deeply rooted habits that they have pursued for decades.\\
While in our work, we argue that small changes that bear some similarity to existing inputs from humans are easier to learn, this is not the sole reason that inputs should resemble similarity to their original inputs. Neglecting this constraint would ultimately lead to proposing the same, ``optimal'' prototypical input for each human. This is not just at odds with the requirement to suggest small changes that are easier to learn, but it also  strongly works against human diversity, which is considered highly valuable.

This work is among the first to discuss the topic of improving human-to-AI interaction through optimization. It contributes as follows: 
\begin{itemize}

\item  Proposing algorithms to optimize single samples of an individual directly in an iterative manner rather than using a separate system to do so. While the latter might be computationally faster, it typically performs some sort of generalization leading to more uniform, less diverse samples. Our approach leads to highly personalized samples, which yields better outcomes in terms of needed time and accuracy, while preserving diversity among humans.
\item  Showing inputs \textit{in combination} with how to create them. That is, we utilize (hand) movement data, highlighting the order and direction of movements rather than exposing a human only to the suggested optimized inputs without any instructions how to create them efficiently.
\item  Conducting an extensive evaluation including a user study, proposing and using metrics that account for discrepancies between suggested inputs and those that are actually created by humans due to innate variation in human movement. 
\end{itemize}

\section{Problem}
This work focuses on object classification. It works on data being the result of human (physical) activity, i.e. hand movements as needed for sketching and writing (and gestures). Each input $X$ by the human to the classifier should be labeled as a specific class $Y$. An input $X$ is a sequence of points $X=((x_i,y_i,I_i))$ ordered in the way the input was created, where $(x_i,y_i)$ are the coordinates of the $i$-th point and $I_i$ is an indicator having one of three values $\{-1,0,1\}$. The value `1' means no line was drawn when moving to this point. `0' indicates a line has been drawn from the prior point $i-1$ to point $i$. The value `-1' only applies for (array) data structures of fixed length, if there are less points then the array length. It indicates that the (array) position is empty. The sequence of points can be split into contiguous line segments, i.e. strokes. A stroke denotes a sequence of connected points, so that no points before and after the sequence (if they exist) are connected to any point of the sequence. We denote as a stroke segment, two connected points of a stroke, ie. $((x_i,y_i,I_i),(x_{i+1},y_{i+1},I_{i+1}))$.
The classifier $C$ processing human inputs is optimized using a known loss $L_{C}$, i.e. typically the cross-entropy loss. The model $C$ is treated as unchangeable, but it can be used in the optimization process. Our human interaction optimization algorithm \emph{O} is provided a sample $X$ with its label $Y$. It computes an optimized input (or proposal) $\hat{X}:=O(X,Y)$ that should improve on $X$ according to one or several objectives. Thus, there is no labeled data available, i.e. there are no optimal proposals, which could be used to train a model. 
We aim to provide some guidance for a user showing how a suggested proposal can be created. That is, we show all operations on how to draw the optimized output $\hat{X}$ based on changing her input $X$. We consider the following alterations of the creation process of the original input: (i) changing the (drawing) direction of a stroke, (ii) changing the position of one or more strokes, (iii) moving a point and (iv) deleting a stroke segment. If a stroke segment is deleted that is not consisting of the last two or first two points of a stroke then a stroke is split into two shorter strokes. 

\subsection{Objectives and Measures}
We consider three main objectives:\\ 
\noindent \textbf{Minimizing time to create inputs}:  While the time to create an optimized sample is easily measured, i.e. in a user study, as an optimization objective, it is difficult to incorporate. In our optimization algorithm, we used a more tangible proxy loss metric to estimate the time to create an optimized sample, i.e. the total distance the hand has to move to create the sketch. We do not include the movement to the start point, but we account for movements for the hand between an end point of a stroke and the starting point of the next stroke. We denote this as effort loss: $$L_{E}(X):=\sum_{i=0}^{|X|-2} \sqrt{(x_i-x_{i+1})^2+(y_i-y_{i+1})^2}$$ 

\noindent \textbf{Minimizing mis-understanding while accounting for human variation}: The amount of wrongly extracted or interpreted information by the AI should be kept as little as possible. That is, optimized inputs should lead to better task performance, ie. higher recognition accuracy, when processed by the AI. Thus, generated sample $\hat{X}$ are created minimizing the classifier loss $L_{C}(\hat{X})$ among other losses. Proposed samples should allow for variance in human behavior. Human behavior is characterized by unintentional variation. For example, a human is not able to reproduce even a single of her own strokes exactly. Since classifiers are knowingly sensitive to small changes in the input as witnessed by adversarial examples, the robustness of optimized samples should be evaluated, e.g. as done for adversarial samples using linear programs \cite{bas16}. We model human variation by creating noisy samples of a proposed input $\hat{X}$. We measure accuracy $Acc_{Noi}$ on these noisy samples. There are multiple approaches to create noisy samples, e.g. using local and global deformations\cite{yu17}. We went for a well-established, easy to comprehend approach: A noisy sample $\hat{X}'$ is created by adding uniform noise to each coordinate. That is for a sequence $\hat{X}$, the noisy sequence $\hat{X}'$ is $$\hat{X}':=\big((\hat{x}_i+\epsilon_{i,0},\hat{y}_i+\epsilon_{i,1},\hat{I}_i)\text{\phantom{a}}|\text{\phantom{a}} (\hat{x}_i,\hat{y}_i,\hat{I}_i) \in \hat{X}\big)$$
Each $\epsilon_{i,j}\in [-r,r]$ is chosen uniformly and independently at random within a fixed range. The range is upper and lower bounded by a constant $r$ that depends on the dataset, e.g. it might be 20 pixels or 1cm.\\

\noindent \textbf{Minimize modifications of original samples}: The proposed samples should bear large similarity to the original inputs. Preserving characteristics of the inputs as much as possible is aligned with the idea that inputs remain comprehensible for other humans or other systems (given that they were so in the first place), diversity is maintained among humans and changes are easy to comprehend and execute for humans. We use two loss objectives depending on what type of alteration is applied to an input.\\
The length-wise difference loss $L_{D}$ captures how much the original and suggested sample differ in visible parts. Length corresponds to the sum of lengths of all visible strokes. That is, only the final outcome is relevant. We do not account for ordering and direction of strokes, which impacts distances to be moved but not the outcome.
The length of visible parts, ie. strokes, of an input $X$ is given by 
\begin{small}
$$V(X):=\sum_{i=0, I_{i+1}\neq1}^{|X|-2} \sqrt{(x_i-x_{i+1})^2+(y_i-y_{i+1})^2} $$
\end{small}
The loss is: $L_{D}(X,\hat{X}):= |V(X)-V(\hat{X})|$ \\
This objective is adequate, if parts of the input are removed. In this case, the original and the modified sample are identical except that one is missing some parts.\\
The point-wise difference loss $L_{P}$ captures the displacement of individual points between the original and suggested sample. It is suitable, when the positions of points are altered, ie. points are moved but their order is kept: 
\begin{small}
$$L_{P}(X,\hat{X}):=\sum_{i=0}^{|X|-1} \sqrt{(x_i-\hat{x}_{i})^2+(y_i-\hat{y}_{i})^2}$$
\end{small}

These three objectives require trade-offs: Keeping changes minimal is at odds with the other objectives, since any change to the input to address the other objectives is non-desirable. Furthermore, minimizing time to create samples is at odds with recognizability. Little time to create implies little information is contained in the outputs, which makes discriminating inputs harder. Thus, what is most preferred -- time or recognizability or minimal changes -- is a decision that is subjective and left to end user. She must state her preferences. We consider two mechanisms a user can state her inclinations: (i) Weighing objectives and (ii) providing constraints. Constraints are ensured not to be violated, while objectives are optimized. But no guarantee can be given beforehand to what extent an objective is fulfilled. To keep matters simple, we shall consider two (primary) objectives, i.e. either minimize ``time'' or maximize ``accuracy'' while constraining the maximal distortion of the original input. Additionally, we enforce that optimized inputs must still be recognizable by the classifier or become recognizable due to the optimization process.


\section{Methodology} 
In our setting, learning requires trial-and-error to identify which alterations of input samples are beneficial due to lack of labeled data. We optimize each input in an iterative fashion, where the strategy to investigate the next option to ``try'' depends on the number of possible solutions. We employ three strategies: gradient descent, a greedy approach and a brute-force approach. Alternatively, one might train a machine learning model to do the optimization of subsequently provided inputs in one forward pass (e.g. \cite{sch20hu,ria18}). A model is likely advantageous if labeled data is available or generalization across samples is helpful or computational load is a concern. But since detailed characteristics of the input should be preserved as much as possible and the diversity of inputs is large, generalization possibilities seem limited. Applying optimization techniques directly on individual samples seems preferable as also confirmed by our experimental evaluation. 

\subsection{Optimizing Individual Samples}
We need to solve a constraint optimization problem encompassing discrete operations, e.g. removal of stroke segments, and continuous operations, e.g. changing coordinates of points. We employ a general framework (Algorithm \ref{alg:gui}) that is adjusted to each alteration operation.

\begin{algorithm}
\caption{Discrete optimization} \label{alg:gui}
\begin{algorithmic}
\STATE \textbf{Input}: Input $X$ of class $Y$, Classifier $C$, Max. distortion $d$, Primary objective $Obj$
\STATE \textbf{Output}: Optimized input $\hat{X}$
\STATE $n_{iter}:=20000$ \COMMENT{number of iterations}
\STATE $\hat{X}:=X$ \COMMENT{Best solution so far}
\STATE $Eval(Z)$:=$\begin{cases} 
       L_D(Z) & Obj=Time \\
       L_C(Z) & Obj=Accuracy \\
   \end{cases} $ 
\FOR{$i$ \textbf{from} 1 \textbf{to} $n_{iter}$}
\STATE $Z:=$ Solution candidate obtained by changing $\hat{X}$ 
\STATE \textbf{If} $L_M(Z)<d\cdot V(X)$ \textbf{then} \COMMENT{Similar to original}
\STATE \quad \textbf{If} $C(Z) =Y \lor C(\hat{X})\neq Y$ \textbf{then} \text{ $\{$ Classified}
\STATE \quad  correctly or no best solution (incl. $X$) has ever been$\}$
\STATE \qquad \textbf{If} $Eval(Z)<Eval(\hat{X})$ \textbf{then} $\hat{X}:=Z$
\ENDFOR
\end{algorithmic}
\end{algorithm}

Algorithm \ref{alg:gui} shows the general outline for our discrete optimization employing (one of the) following operations: removal of strokes, reversing stroke direction or changing stroke order. The maximal allowed deviation $d$ is specified by the user as well as the objective $obj$, which is either time or accuracy. The algorithm creates a new solution candidate $Z$ in each iteration using the considered operation. A solution candidate $Z$ is checked if it fulfills the following two constraints: (i) It does not deviate too much from the original input; (ii) it is classified correctly or no best solution $\hat{X}$ has been classified correctly so far. \footnote{Note, this includes the original input $X$ due to initialization of $\hat{X}:=X$.}  If a solution candidate $Z$ fulfills both constraints and it has lower loss value according to the user determined objective $obj$, the best solution $\hat{X}$ is set to the solution candidate $Z$.
Solution candidates are generated for each operation as follows: 
For removal of visible parts, we consider all stroke segments $s_i:=((x_i,y_i,I_i),(x_{i+1},y_{i+1},I_{i+1}))$. A solution candidate $Z$ based on $\hat{X}$ is obtained by removing point $i$ from $\hat{X}$, ie. $Z:=X\setminus s_i$. In case, point $i+1$ is the last point of a stroke, it is also removed. Otherwise, additionally, we set $I_{i+1}=1$ to indicate the start of a new stroke. The order in which the solution candidates are created is important. That is, it matters which stroke segments are removed first. A natural choice (denoted as CL) is to select stroke segments with smallest classifier loss $L_C$ as being removed first. The idea being that segments being only weakly indicative of the given class $Y$ can likely be removed without causing a mis-classification. Increasing the number of strokes due to removal of stroke segments is not desirable. Therefore, we consider a variant (CE), where only stroke segments at the beginning or end of a stroke are removed.
For comparison, we also consider removal in reverse order(RO) in which points are created and in the same order(SO). The motivation being that this procedure constantly removes one stroke after the other without splitting strokes. Furthermore, humans might first draw the most important high level outline that might help in distinguishing objects and add more details over time. 


For changing the direction of strokes and changing the order of strokes, we choose a solution candidate randomly. That is, for changing direction of a stroke, we flip the direction of a randomly chosen stroke of $\hat{X}$ by reversing the sequence of points of the stroke. To change the order of strokes, we perform a cut-and-paste operation. That is, we choose a random sequence consisting of subsequent strokes, remove it and insert it either after a random stroke or at the very beginning of the sequence. We also consider doing both that is for each cut-and-paste operation, we flip randomly the direction with 50\% probability. This randomized approach is essentially equally to trying all options given the number of strokes is small.

For continuous optimization, we use gradient descent with gradients obtained from the classifier. That is, we maintain a solution candidate $Z$ that is initialized with the original input $X$. It is updated in each iteration using a gradient descent step treating the classifier weights as fixed and the input $X$ as variable. Otherwise the procedure is identical to Algorithm \ref{alg:gui}. The loss function $L_{Tot}$ is a weighted combination of the classifier loss $L_C$, the effort loss $L_M$ and the point-wise difference loss $L_{P}$:  
\begin{small}
$$L_{Tot}(Z):= \beta_{C}L_C(Z)+\beta_{P}L_{P}(Z)+\beta_{E}L_{E}(Z)$$  
\end{small}

\section{Evaluation} 
\noindent\textbf{Datasets}: We used two datasets. 1 Mio. samples of the QuickDraw \cite{ha17} dataset distributed equally among the first 30 classes. It consists of human sketches of an object given its name. The data was created primarily using mouse and touch devices. The second dataset consists of musical notes drawn using a pen\cite{cal14}. It consists of 32 classes but only of 15200 samples. We padded or stripped sequences to be of fixed length, ie. 104 points. For evaluation, we optimized 256 samples for each class of the QuickDraw dataset that were not used for training, yielding 7680 test samples. For the Homus dataset we used 20\% of the data corresponding to 3040 samples. Note that our optimization algorithms do not require any training data. They optimize test samples directly.\\ 
\noindent\textbf{Models}: For classifier $C$, we trained three instances of two identical classifier architectures for each dataset yielding 12 classifiers. For each instance, we ran our optimization algorithms. We used PyTorch 1.6.0 and trained on 2080 TI Nvidia GPUs. Our ``LSTM'' architecture is made of 3 Conv1D, 2 LSTM and 2 dense layers with dropout. The ``Conv1D'' network consists of 3 Conv1D layers with stride 2 and a dense layer. For comparing to prior work, we also implemented an architecture based on \cite{yu17}. Details on architectures can be found in the supplementary material. If not stated differently we used $d=20\%$, ie. we allowed at most a distortion of 20\% for deletion and moving points by at most 20\%. 

\noindent\textbf{Procedure}: We conduct: i) a user study, where humans have to redraw original and optimized samples, ii) a qualitative evaluation, illustrating created samples, and iii) a quantitative evaluation, discussing the impact of parameters and comparing various approaches.  Results for all combinations of models and datasets were very similar in nature. Therefore, we focus on just one scenario, i.e. the LSTM-network on the Quickdraw dataset, results for other setups are summarized in the end and detaisl can be found in the supplementary material.


\subsection{Qualitative Evaluation}
We discuss optimized samples for each alteration operation separately. 
\begin{figure}[!ht]
	\centering \includegraphics[width=0.47\textwidth]{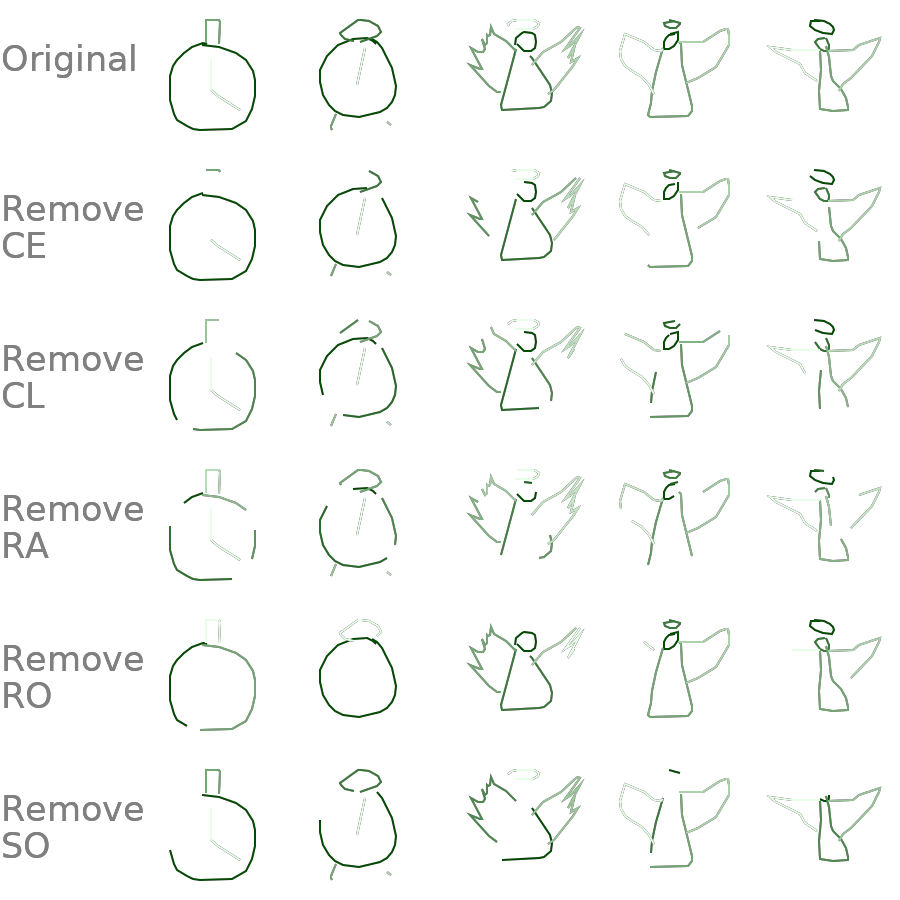}
	\caption{Original and generated samples for removal, minimizing creation effort - more samples in supplementary material} \label{fig:clmb}
\end{figure} 
As seen in Figure \ref{fig:clmb} methods preferring removal of segments in the same order as creation (SO) or reverse order(RO) tend to remove entire strokes. This can lead to unnatural sketches, e.g. angles without heads. Random removal(RA) and classifier loss based ordering (CL) increases the number of strokes, which might be undesirable, when it comes to reproducing the optimized sample. Removing only end points based on classifier loss (CE) tends to produce samples that are well-recognizable without artifacts and it does not split strokes. 

\begin{figure*}[!ht]
	\centering \includegraphics[width=0.7\textwidth]{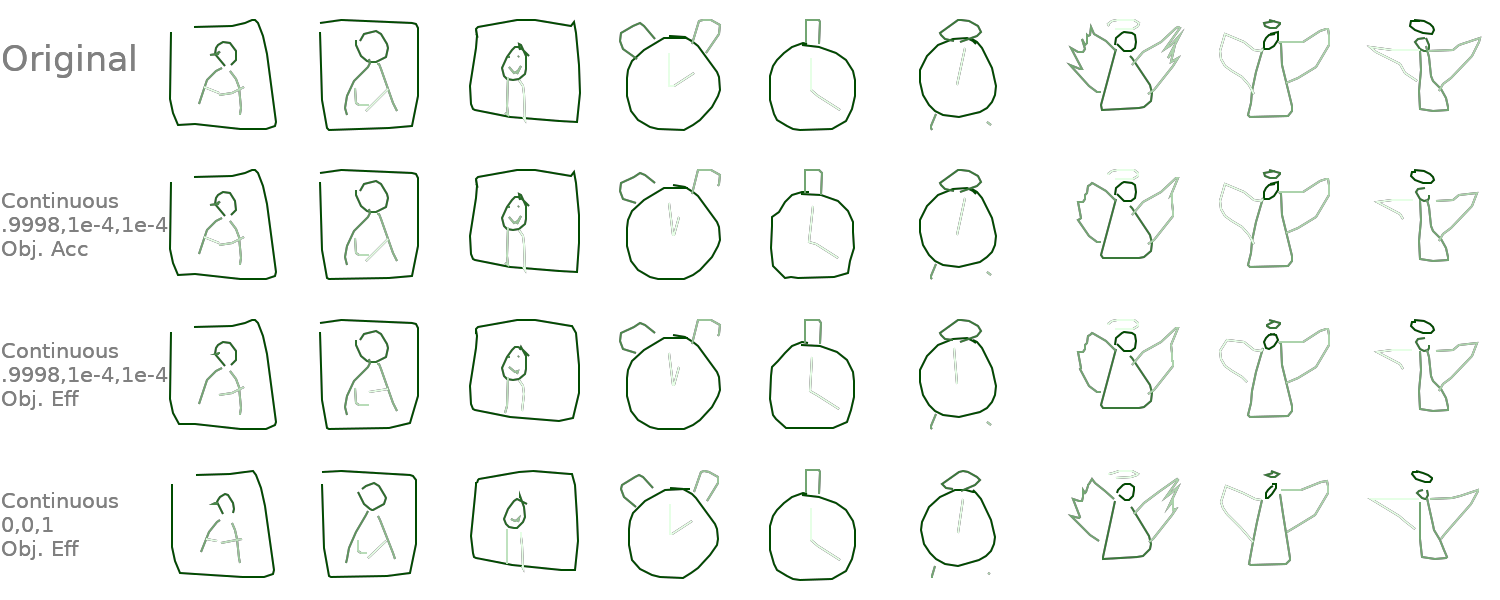}
	\caption{Original and generated samples for continuous optimization for various $\beta_{C}$, $\beta_{P}$ and $\beta_{E}$} \label{fig:cont}
\end{figure*}
Figure \ref{fig:cont} shows outcomes for continuous optimization. Changes appear more subtle than for removal, in particular, when optimizing for accuracy (second row). Continuous optimization tends to shorten strokes, straighten them (best seen for wings of the first angle) and it might also rotate them -- as done for clock hands. When only optimizing for effort (last row), which should lead to largest distortion, changes seem to be the least noticeable. Our quantitative analysis shows that this is not the case. Samples get scaled entirely in a more uniform manner, making changes harder to spot (best seen for third image from the left in the last row). 


Samples for altering order and direction of strokes are in the supplementary material.

\subsection{Quantitative}
We use priorly described metrics related to (i) the classifier's capability to recognize samples ($Acc$(uracy), $Acc_{Noi(se)}$), (ii) the effort of creating a sample $L_{E}$ and (iii)  the distortion of the original sample $L_{D}, L_{P}$. To compute $Acc_{Noi}$, we create for each input 10 noisy samples,  where each $\epsilon_{i,j}\in [-10,10]$. For the Quickdraw dataset, where points are within a range of [0,255] this means that the maximal distance due to the addition of $\epsilon_{i,j}$ to each coordinate between two points is about 28 pixels or 10\% of the canvas used for sketching.  We report the accuracy on all created samples, i.e. for the Quickdraw dataset with 7680 test samples, we report the accuracy of 76800 samples.\\


Table \ref{tab:cres} shows all metrics for continuous optimization varying loss weights $\beta$. All settings improve upon the original in terms of accuracy. This is expected given the constraint that a sample is no modification is done that changes a correctly classified sample into an incorrectly classified one. When optimizing for effort, accuracy gains of optimized samples vary. For noisy samples, accuracy can even be lower than for the original. Reduced samples contain less (redundant) information for classification than the original ones, making them somewhat more sensitive to noise. If optimizing effort loss ($\beta_{E}=1$) only, effort loss is not lowest among all options. Having a classifier loss $\beta_C>0$, not only strongly improves accuracy, but interestingly also leads to lowest effort loss. Without a  classifier loss, all parts of the sketch are altered irrespective of whether they are relevant for classification. Thus, significant increase in loss is occurred due to small movements of highly relevant points for classification. This is largely avoided using a classifier loss.

\begin{table}[!htb]
    \setlength\tabcolsep{4.5pt}
	\centering
	\scriptsize
	\begin{tabular}{|c| c|c|c| c|	c|	} \hline 
	 Obj &$\beta_{C}$, $\beta_{P}$, $\beta_{E}$ & $Acc$& $Acc_{Noise}$ & $L_{E}$ & $L_{P}$ \\ \hline 
\multicolumn{2}{|l|}{Original}& 0.897\text{\tiny{$\pm$0.02}}&0.894\text{\tiny{$\pm$0.02}}&1527.4\text{\tiny{$\pm$161.4}}&0.0\text{\tiny{$\pm$0.0}} \\ \hline\hline
\multirow{4}{*}{\centering Acc}&1.0,0,0&0.929\text{\tiny{$\pm$0.04}}&0.923\text{\tiny{$\pm$0.03}}&1539.6\text{\tiny{$\pm$158.8}}&\textbf{10.8}\text{\tiny{$\pm$11.5}}\\ \cline{2-6}
&0,0,1.0&0.904\text{\tiny{$\pm$0.02}}&0.9\text{\tiny{$\pm$0.02}}&1485.8\text{\tiny{$\pm$162.8}}&39.3\text{\tiny{$\pm$9.1}}\\ \cline{2-6}
&.9999,0,.0001&0.963\text{\tiny{$\pm$0.01}}&0.954\text{\tiny{$\pm$0.01}}&1461.9\text{\tiny{$\pm$160.1}}&63.0\text{\tiny{$\pm$10.2}}\\ \cline{2-6}
&.9998,.0001,.0001&\textbf{0.966}\text{\tiny{$\pm$0.01}}&\textbf{0.958}\text{\tiny{$\pm$0.01}}&\textbf{1446.9}\text{\tiny{$\pm$159.8}}&78.2\text{\tiny{$\pm$16.9}}\\ \hline\hline
\multirow{4}{*}{\centering Eff}&1.0,0,0&0.962\text{\tiny{$\pm$0.01}}&0.927\text{\tiny{$\pm$0.01}}&1525.3\text{\tiny{$\pm$162.3}}&\textbf{1.94}\text{\tiny{$\pm$1.1}}\\ \cline{2-6}
&0,0,1.0&0.903\text{\tiny{$\pm$0.02}}&0.89\text{\tiny{$\pm$0.02}}&\textbf{1407.1}\text{\tiny{$\pm$152.3}}&113.2\text{\tiny{$\pm$13.4}}\\ \cline{2-6}
&.9999,0,.0001&0.963\text{\tiny{$\pm$0.01}}&0.934\text{\tiny{$\pm$0.01}}&1422.0\text{\tiny{$\pm$157.9}}&100.1\text{\tiny{$\pm$9.1}}\\ \cline{2-6}
&.9998,.0001,.0001&\textbf{0.966}\text{\tiny{$\pm$0.01}}&\textbf{0.94}\text{\tiny{$\pm$0.01}}&1417.1\text{\tiny{$\pm$160.4}}&107.6\text{\tiny{$\pm$24.4}}\\ \hline
\end{tabular}
	\caption{Results varying loss term weights $\beta_{CL}$, $\beta_{RE}$, $\beta_{EF}$}	\label{tab:cres}	 
\end{table}

\begin{table}[!htb]
    \setlength\tabcolsep{4.5pt}
	\centering
\scriptsize
	\begin{tabular}{|c|c|c|c|c|c|} \hline
	 Obj & Meth. & $Acc$ & $Acc_{Noi}$ & $L_{E}$ & $L_{D}$ \\ \hline 
\multicolumn{2}{|l|}{Original}& 0.897\text{\tiny{$\pm$0.02}}&0.895\text{\tiny{$\pm$0.02}}&1527.4\text{\tiny{$\pm$161.4}}&0.0\text{\tiny{$\pm$0.0}} \\ \hline\hline
\multirow{5}{*}{\centering Acc}&CL&\textbf{0.971}\text{\tiny{$\pm$0.0}}&\textbf{0.967}\text{\tiny{$\pm$0.01}}&1444.2\text{\tiny{$\pm$157.1}}&180.7\text{\tiny{$\pm$15.7}}\\ \cline{2-6}
&CE&0.956\text{\tiny{$\pm$0.01}}&0.953\text{\tiny{$\pm$0.01}}&\textbf{1395.4}\text{\tiny{$\pm$151.7}}&123.7\text{\tiny{$\pm$11.3}}\\ \cline{2-6}
&RA&0.919\text{\tiny{$\pm$0.01}}&0.916\text{\tiny{$\pm$0.01}}&1503.2\text{\tiny{$\pm$159.0}}&60.9\text{\tiny{$\pm$8.7}}\\ \cline{2-6}
&RO&0.908\text{\tiny{$\pm$0.01}}&0.906\text{\tiny{$\pm$0.01}}&1497.7\text{\tiny{$\pm$160.7}}&\textbf{24.6}\text{\tiny{$\pm$4.5}}\\ \cline{2-6}
&SO&0.911\text{\tiny{$\pm$0.01}}&0.909\text{\tiny{$\pm$0.02}}&1499.0\text{\tiny{$\pm$160.0}}&29.3\text{\tiny{$\pm$3.8}}\\ \hline\hline
\multirow{5}{*}{\centering Eff}&CL&\textbf{0.961}\text{\tiny{$\pm$0.01}}&\textbf{0.955}\text{\tiny{$\pm$0.01}}&1346.2\text{\tiny{$\pm$141.0}}&220.5\text{\tiny{$\pm$20.0}}\\ \cline{2-6}
&CE&0.947\text{\tiny{$\pm$0.01}}&0.941\text{\tiny{$\pm$0.01}}&1257.3\text{\tiny{$\pm$131.0}}&219.3\text{\tiny{$\pm$19.8}}\\ \cline{2-6}
&RA&0.914\text{\tiny{$\pm$0.01}}&0.881\text{\tiny{$\pm$0.02}}&1381.8\text{\tiny{$\pm$143.4}}&220.7\text{\tiny{$\pm$19.8}}\\ \cline{2-6}
&RO&0.912\text{\tiny{$\pm$0.01}}&0.872\text{\tiny{$\pm$0.02}}&\textbf{1191.1}\text{\tiny{$\pm$122.9}}&\textbf{219.1}\text{\tiny{$\pm$19.7}}\\ \cline{2-6}
&SO&0.912\text{\tiny{$\pm$0.01}}&0.868\text{\tiny{$\pm$0.02}}&1288.6\text{\tiny{$\pm$140.4}}&220.2\text{\tiny{$\pm$20.0}}\\ \hline
\end{tabular}
	\caption{Results for removal}	\label{tab:gre}	 
\end{table}

The results for all removal strategies (Table \ref{tab:gre}) indicate that abandoning irrelevant stroke parts yields improvements for both effort and accuracy at the same time. Using the classifier loss (CL or CE) for ordering removals yields best results in terms of accuracy. When optimizing for effort, only these methods also achieve much better noisy accuracy $Acc_{Noi}$ than the original samples. Both also do well in terms of effort. When being allowed to split strokes (CL) accuracy is larger than for removing parts at the end of strokes CE, but these gains come at the expense of having more strokes. Removing samples in reverse order (RO) or in sorted order (SO) gives lowest effort loss. Removing entire strokes from the beginning (or end) yields benefits, since we do not account for moving to the first point or from the last point to some starting point and there is often a significant distance between the end point of one stroke and start point of the next stroke. This distance is also gained when removing entire strokes. In contrast, when removing strokes (segments) in the middle, a transition between strokes remains.

\begin{table}[!htb]
    \setlength\tabcolsep{4.5pt}
	\centering
	\scriptsize
	\begin{tabular}{|c| c|c|c|} \hline 
	    Method & Original Acc. & $\Delta$ if keep 50\% & $\Delta$ if keep 25\% \\ \hline
	    DQSN & 0.92 &  -0.12 & -0.27\\ \hline
	    GDSA& 0.92 & -0.06 & -0.20\\ \hline
	    CE(This paper) & 0.95 &  -0.02 & -0.16\\ \hline
	    CL(This paper) & 0.95 &  \textbf{0.01} & \textbf{-0.14}\\ \hline
	 	\end{tabular}		
	\caption{Accuracy reduction for Sketch-a-Net architecture when reducing visible elements; bold shows best  }\label{tab:cvp}	
\end{table}

Table \ref{tab:cvp} shows that the proposed optimization procedure if only a fixed percentage of average stroke segments of a category is kept as proposed and described for GDSA in \cite{muh19} for a smaller subset of QuickDraw. In this setup, removal takes place even if it leads to erroneous classification. Our classifier guided methods CE and CL achieve higher accuracy than prior work (DQSN\cite{zhou18vid} and GDSA) which is based on training a model using reinforcement learning significantly. We attribute this to the fact that we optimize samples individually in an iterative manner.

As shown in Table \ref{tab:percomb} both permuting strokes (P) and reversing directions (R) or doing both (B) yield significant gains in terms of accuracy and also effort.

\begin{table}[!htb]
    \setlength\tabcolsep{4.5pt}
	\centering
	\scriptsize 
	\begin{tabular}{|c|c|c|c| c|c|} \hline
	 Obj & Meth. & $Acc$ & $Acc_{Noi}$ & $L_{E}$ & $L_{D}$ \\ \hline 

\multicolumn{2}{|l|}{Original}& 0.897\text{\tiny{$\pm$0.02}}&0.895\text{\tiny{$\pm$0.02}}&1527.4\text{\tiny{$\pm$161.4}}&0.0\text{\tiny{$\pm$0.0}} \\ \hline
\multirow{3}{*}{\centering Acc}&P&0.952\text{\tiny{$\pm$0.01}}&0.948\text{\tiny{$\pm$0.01}}&1595.6\text{\tiny{$\pm$162.1}}&0.096\text{\tiny{$\pm$0.07}}\\ \cline{2-6}
&R&0.951\text{\tiny{$\pm$0.01}}&0.948\text{\tiny{$\pm$0.01}}&\textbf{1551.8}\text{\tiny{$\pm$159.5}}&\textbf{0.06}\text{\tiny{$\pm$0.07}}\\ \cline{2-6}
&B&\textbf{0.961}\text{\tiny{$\pm$0.01}}&\textbf{0.959}\text{\tiny{$\pm$0.01}}&1580.1\text{\tiny{$\pm$159.6}}&0.094\text{\tiny{$\pm$0.04}}\\ \hline
\multirow{3}{*}{\centering Eff}&P&0.911\text{\tiny{$\pm$0.02}}&\textbf{0.892}\text{\tiny{$\pm$0.02}}&1393.9\text{\tiny{$\pm$135.7}}&0.179\text{\tiny{$\pm$0.09}}\\ \cline{2-6}
&R&\textbf{0.915}\text{\tiny{$\pm$0.01}}&0.888\text{\tiny{$\pm$0.01}}&1306.6\text{\tiny{$\pm$120.5}}&0.193\text{\tiny{$\pm$0.1}}\\ \cline{2-6}
&B&0.913\text{\tiny{$\pm$0.01}}&0.88\text{\tiny{$\pm$0.02}}&\textbf{1295.2}\text{\tiny{$\pm$117.9}}&\textbf{0.096}\text{\tiny{$\pm$0.1}}\\ \hline

\end{tabular}		
	\caption{Results for permuting strokes(R), reversing direction (R) and doing both (B)}	\label{tab:percomb}	
\end{table}

\begin{table}[!htb]
    \setlength\tabcolsep{4.5pt}
	\centering
	\scriptsize
	\begin{tabular}{|c|c|c|c| c|	c|c|c|	} \hline
	Obj& Met. & $Acc$&$Acc_{Noi}$& $L_{E}$ & $L_{D}$ & $L_{P}$ \\ \hline 
\multicolumn{2}{|l|}{Original}& 0.897\text{\tiny{$\pm$0.02}}&0.895\text{\tiny{$\pm$0.02}}&1527.4\text{\tiny{$\pm$161.4}}&0.0\text{\tiny{$\pm$0.0}}&0.0\text{\tiny{$\pm$0.0}} \\ \hline\hline
\multirow{3}{*}{\centering Acc}&B-C&0.976\text{\tiny{$\pm$0.0}}&0.972\text{\tiny{$\pm$0.0}}&1518.4\text{\tiny{$\pm$155.5}}&\textbf{60.1}\text{\tiny{$\pm$13.7}}&69.1\text{\tiny{$\pm$12.0}}\\ \cline{2-7}
&C-D&0.975\text{\tiny{$\pm$0.0}}&0.972\text{\tiny{$\pm$0.01}}&\textbf{1388.4}\text{\tiny{$\pm$150.0}}&167.2\text{\tiny{$\pm$25.6}}&107.2\text{\tiny{$\pm$18.5}}\\ \cline{2-7}
&D-B&\textbf{0.98}\text{\tiny{$\pm$0.01}}&\textbf{0.979}\text{\tiny{$\pm$0.01}}&1486.4\text{\tiny{$\pm$161.8}}&152.2\text{\tiny{$\pm$32.2}}&\textbf{0.0}\text{\tiny{$\pm$0.0}}\\ \hline\hline
\multirow{3}{*}{\centering Eff}&B-C&0.961\text{\tiny{$\pm$0.01}}&0.925\text{\tiny{$\pm$0.01}}&1203.3\text{\tiny{$\pm$120.0}}&\textbf{101.4}\text{\tiny{$\pm$22.2}}&131.1\text{\tiny{$\pm$14.1}}\\ \cline{2-7}
&C-D&\textbf{0.973}\text{\tiny{$\pm$0.0}}&\textbf{0.964}\text{\tiny{$\pm$0.01}}&1281.7\text{\tiny{$\pm$132.8}}&228.7\text{\tiny{$\pm$21.5}}&121.3\text{\tiny{$\pm$18.5}}\\ \cline{2-7}
&D-B&0.958\text{\tiny{$\pm$0.01}}&0.911\text{\tiny{$\pm$0.02}}&\textbf{1086.1}\text{\tiny{$\pm$94.2}}&219.9\text{\tiny{$\pm$19.2}}&\textbf{0.0}\text{\tiny{$\pm$0.0}}\\ \hline
\end{tabular}
	\caption{Results for applying multiple methods sequentially; (C)ontinuous point movement, (B) Reverse and permute, (D)eletion of parts}	\label{tab:comb}	 
\end{table}

We also applied two and more  methods sequentially (Table \ref{tab:comb}). For continuous optimization of points $C$ we used $(\beta_{C},\beta_{P},\beta_{E})= (.9998,.0001,.0001)$. Applying multiple methods gives some more improvement. That is, both the maximum accuracy and minimum effort loss are lower, when applying multiple methods. We also investigated different orderings, eg. B-C instead of C-B as well as performing multiple applications simulating an interwoven optimization. For example, B-C-B-C with reduced iterations for each method. This leads to some additional improvements. \\
\textbf{Other networks and datasets:} We found that qualitatively all results were identical, meaning that if there was a clear improvement for optimized samples for one dataset and one network type this also held for others. But gains could vary per dataset, network and operation considered. For example, the highest accuracy gains relative to the original was achieved for Conv1D on Quickdraw (13.4\%) compared to 8.3\% for LSTM on QuickDraw.



\section{User study} 
\begin{figure}[!ht]
	\centering \includegraphics[width=0.47\textwidth]{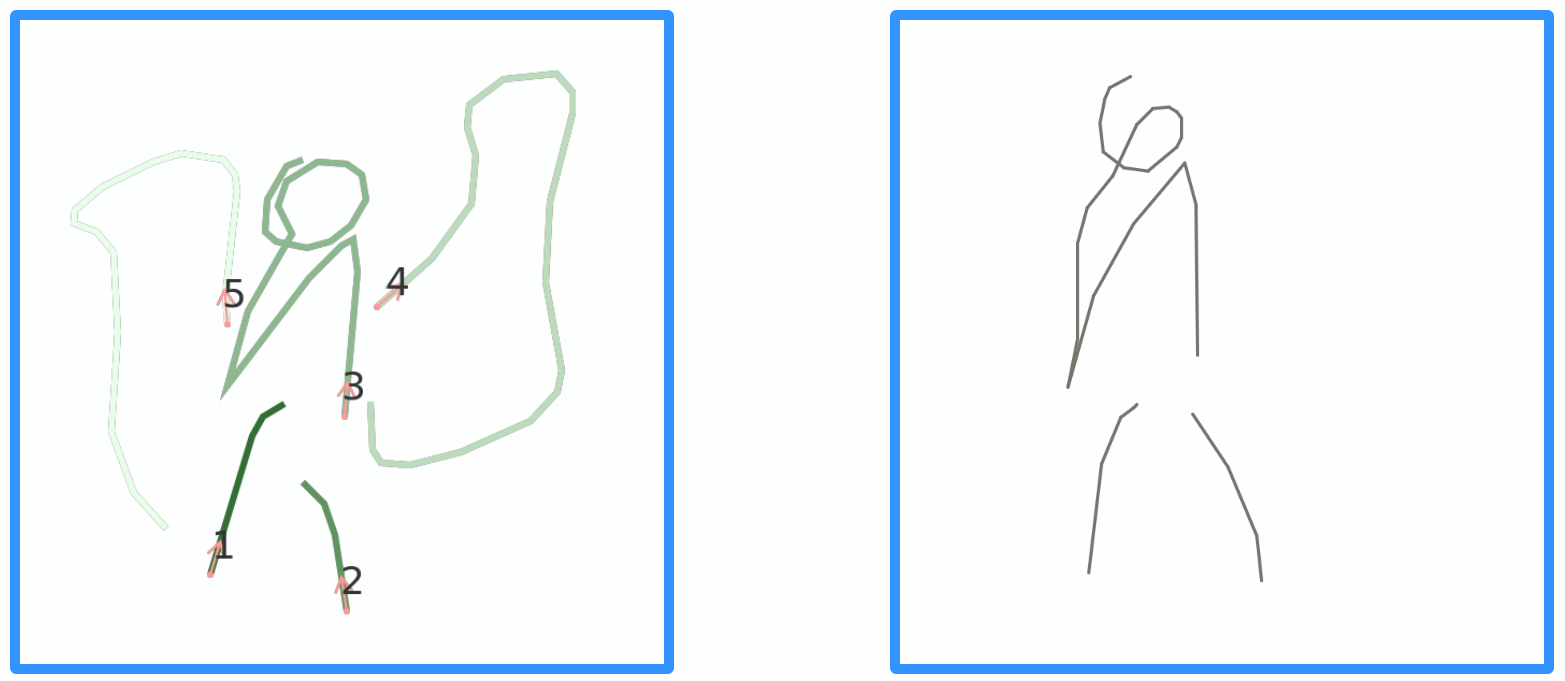}
	\caption{During the user study participants are shown a sketch with numbered strokes and stroke start indicated (left panel). They should reproduce it (right)} \label{fig:gred}
\end{figure}

We conducted an experiment to assess, if optimized samples can be reproduced by humans and if these reproductions indeed yield gains according to the specified objective. While the prior numerical investigation is highly suggestive, optimized samples might be unnatural for humans. Thus, reproductions of those samples might take longer and deviate more strongly from the proposal than non-optimized samples, making a user study necessary.
We used generated samples for method ``D-B'' (Table \ref{tab:comb}) optimized towards accuracy for the QuickDraw dataset. The overall pool of sketches consisted of 10 original samples per class, where each sample consisted of up to 7 strokes to ensure good readability of instructions, ie. numbering and arrows. Since we are particularly interested in the capability, whether errors in interaction can be mitigated, we chose 5 (of the 10) original samples per class that were misclassified. 
Each participant had to copy an optimized version of a human input and the original version for five randomly selected sketches, yielding 10 sketches per user - see Figure \ref{fig:gred}. It was decided randomly, whether a user was first presented the optimized or original version. Users were advised to use the same number of strokes and draw them in the order and direction as indicated by the numbering and arrows.
We recruited 200 English speaking participants on Amazon Mechanical Turk. We removed reproduced sketches, that did not match the instructed  number of strokes or took more than 60s to create sketch or that had only the original or the corresponding optimized sample drawn adequately, i.e. within 60s and with the correct number of strokes. 
The (LSTM) classifier had an accuracy of 54\% on sketches resembling the original and 68\% on sketches based on the optimized sample. The differences are statistically significant using a t-test, yielding $p<0.02$. Participants took on average 23.2s to (re)sketch an original sample. They were 1.7s faster for optimized samples (though only with $p=0.21$). Note that we used samples optimized towards accuracy not effort. Still, even those samples have (mostly) less visible strokes ($L_D$), while overall hand movements are typically similar to original samples ($L_E$) - see Table \ref{tab:comb}.

\section{Related Work}\label{sec:related work} 
\emph{Human-AI interaction:} \cite{rze18} summarized the effects of various user and AI system characteristics in general, while \cite{mar19} focused on digital AI assistants. Interaction between AI and users has also been studied \cite{ame19,jan19,ban19,car19,mar19,noc19,gho19a,shn20} in various contexts such as social robots \cite{mar19,noc19}. The primary focus has been on desirable AI behavior, eg. empathy, or strategies how AI can adapt to user behavior(\cite{gho19a,car19} with few exceptions. \cite{ban19} explicitly investigated how users can alter the behavior, i.e. override decisions of the AI, by understanding the error boundary of a classifier, while \cite{shn20} provides general guidelines on human-centered AI. Closest to our work is \cite{sch20hu}. \cite{sch20hu} introduced a human-to-AI coach based on an auto-encoder that given a picture of a digit outputs a digit that has lower classifier loss and potentially consists of less pixels. In contrast, we optimize samples individually in an iterative manner, also provide instructions on how to create samples and we are the first to evaluate on actual users. Our work also uses more complex datasets that are commonly studied in other contexts, e.g. see \cite{xu20} for a survey on machine learning and human sketches.   Abstracting sketches using removal of stroke segments and entire strokes, while preserving semantics was studied in \cite{ria18,muh19}. That is, an agent learns to select strokes that are relevant for a classifier to maintain the correct class. From the perspective of this paper, this is similar to neglecting all constraints and focusing on ``time'' with narrowing down on just one option for abstracting: Removal. In this paper, we also consider a gradual movement of points. \cite{ria18,muh19} used reinforcement learning while having correct classification as a reward (rather than as an objective). The implementation using reinforcement learning also differs from our approach improving individual samples directly. Combination of both approaches might lead to better outcomes that is using RL with a planning and simulation capability. \cite{li19} used GANs to complete artificially corrupted sketches, ie. through occlusion. They achieved high-quality results comparable to other methods such as image inpainting.\\
\noindent\emph{Explainability:} This paper has strong ties to (personalized) explanations \cite{sch19,gui18} and explanations in the field of human-AI interaction \cite{hoi19}. Counterfactual explanations seek to identify a modification of the input to obtain another class \cite{dhu18,goy19}. \cite{dhu18} aims to identify minimal changes to digits on a pixel level using perturbations. Thus, in contrast to our work, they focus on mis-classified samples. Moreover, the suggested changes commonly involve adding or removing multiple pixels distributed across the digit, which seems infeasible for humans, since they are not able to reproduce digits on that level of detail. \cite{goy19} combine two images, the image to change and an image from a class the image should be changed to. 

 

\section{Discussion and Conclusions} \label{sec:conc}
Humans will interact more and more with AI. This paper provided first steps towards improving this interaction by showing how human inputs to an AI can be optimized. Our results indicate that optimized samples can be created faster and lead to less mis-classifications, while still bearing similarity to the original input. We chose to optimize samples individually, which allows to personalize to a very high degree. We believe that an approach using meta-learning or reinforcement learning with planning could lead to even better results. Our optimized samples were also accompanied by instructions on how to create them efficiently. While our human study confirmed that reconstructing proposed samples leads to saving in time and reduces mis-classifications, more exploration of the field of human to AI interaction is needed: improve interaction on a semantic level as needed for interaction with chatbots beyond making chatbots more human \cite{cha19,cie19}, consider other recognition problems such as speech\cite{zhan18}, perform a joint optimization of human inputs and AI models, eg. interactive modeling\cite{war01}, derive optimization algorithms that use inputs of a human to provide general rules as feedback, assess additional concerns such as acceptance of technology by humans\cite{ven03}. We hope that the community will pick up on these questions to foster seamless use of AI and reduce risks due to miscommunication.


\bibliography{references}
\bibliographystyle{aaai} 
\end{document}